\documentclass[11pt,twoside]{article}
\usepackage{cal05,graphicx}

\contact{Martin K\"ummel}
\email{mkuemmel@eso.org}

\begin{document}

\title{Slitless spectroscopy with the Advanced Camera for Surveys}
\titlemark{Slitless spectroscopy with ACS}

\author{M.\ K\"ummel, S.S.\ Larsen and J.R.\ Walsh}
\affil{Space Telescope - European Coordinating Facility,
Karl-Schwarzschild-Str. 2, D-85748 Garching b. M\"unchen, Germany}

% \author{}
% \affil{}

\paindex{K\"ummel, M.}
\aindex{Larsen, S. S.}
\aindex{Walsh, J.R.}

%-----------------------------------------------------------------------
%                     Author list for page header
%-----------------------------------------------------------------------
% Please supply a list of author last names for the page header. in
% one of these formats:
%
% EXAMPLES:
% \authormark{LASTNAME}
% \authormark{LASTNAME1 \& LASTNAME2}
% \authormark{LASTNAME1, LASTNAME2, ... \& LASTNAMEn}
% \authormark{LASTNAME et al.}
%
% Use the "et al." form in the case of seven or more authors, or if
% the preferred form is too long to fit in the header.

\authormark{}

% The abstract is entered in a LaTeX "environment", designated with paired
% \begin{abstract} -- \end{abstract} commands. Other environments are
% identified by the name in the curly braces.

% Poster authors ONLY may omit the abstract in order to gain a little
% more page space for the text of the poster.

\begin{abstract}
The Advanced Camera for Surveys (ACS) enables low resolution
slitless spectroscopic imaging in the three channels. The most-used
modes are grism imaging with the WFC and the HRC at a resolution of
$40$ and 24 \AA/pixel, respectively. In the far UV there are two prisms
for the SBC and a prism for the HRC in the near-UV. 
An overview of the slitless spectroscopic modes of the ACS is
presented together with the advantages of slitless spectroscopy
from space. The methods and strategies developed to establish and maintain 
the wavelength and flux calibration for the different channels are outlined. 
Since many slitless spectra are recorded on one deep exposure, pipeline 
science quality extraction of spectra is a necessity. To reduce ACS slitless
data, the aXe spectral extraction software has been developed at the ST-ECF. 
aXe was designed to extract large numbers of ACS slitless spectra
in an unsupervised way based on an input catalogue derived from a companion
direct image. In order to handle dithered slitless spectra, drizzle, 
well-known in the imaging domain, has been applied. For ACS grism 
images, the aXedrizzle technique resamples 2D spectra from individual 
images to deep, rectified images before extracting the 1D spectra. aXe 
also provides tools for visual assessment of the extracted spectra and 
examples are presented. 
\end{abstract}

%-----------------------------------------------------------------------
%			Subject Index keywords
%-----------------------------------------------------------------------
% Enter up to 6 keywords describing your paper.  These will NOT be
% printed as part of your paper; however, they will be used to
% generate the subject index for the proceedings.  There is no
% standard list; however, you can consult the indices for past Calibration
% Workshop Proceedings. 


%-----------------------------------------------------------------------
%			      Main Body
%-----------------------------------------------------------------------
% Place the text for the main body of the paper here.  You should use
% the \section command to label the various sections; use of
% \subsection is optional.  Significant words in section titles should
% be capitalized.  Sections and subsections will be numbered
% automatically. 
\section{Grisms and prisms on the ACS}
The ACS has three channels, the Wide Field Channel (WFC), the High resolution
Channel (HRC) and the Solar Blind Channel (SBC), and each channel is capable
of delivering slitless spectroscopic images by inserting a grism or prism into
the optical beam. The five different combinations of ACS channel and
dispersing element offer low resolution spectroscopy from the UV to the far-red
wavelength regime. Table \ref{spec_modes} lists important parameters such as
spectral resolution, wavelength range and field of view (FoV) for all slitless
modes of ACS.\

\begin{figure}
\epsscale{0.30}
%\plotone{kuemmelF1.ps}
\includegraphics[angle=-90,width=.8\textwidth]{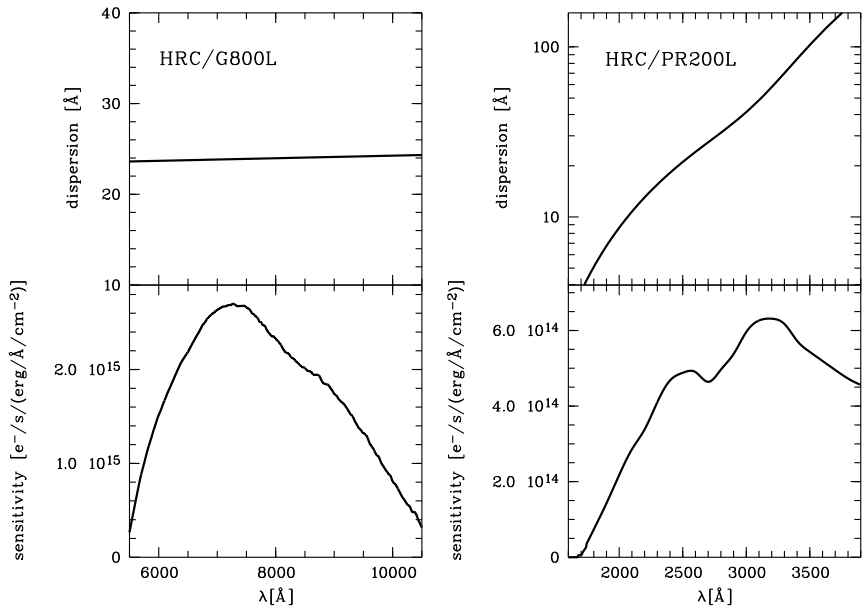}
\caption{A comparison between dispersion and sensitivity of the HRC
grism G800L (left) and prism PR200L (right).}
\label{example-fig-1}
\end{figure}

\begin{table}
\begin{tabular}{cccccc}
\tableline
\tableline
Channel & Disperser & Wav. Range & Resolution & Pixel Size & FOV \\
        &           &      [\AA]       & [\AA/pixel]& [mas/pixel]& [arcsecond]\\
\tableline
WFC     & G800L     & $5500-10500$     &     $38.5$   & $50\times 50$& $202\times 202$\\
HRC     & G800L     & $5500-10500$     &     $23.5$   & $28\times 24$& $29\times 26$\\
HRC     & PR200L     & $1700-3900$     & 20[@2500\AA] & $28\times 24$& $29\times 26$\\
SBC     & PR130L     & $1250-1800$     & 7[@1500\AA]   & $34\times 30$& $35\times 31$\\
HRC     & PR110L     & $1150-1800$     & 10[@1500\AA] & $34\times 30$& $35\times 31$\\
\tableline
\end{tabular}
\caption{The spectroscopic modes of the ACS.}
\label{spec_modes}
\end{table}
Figure \ref{example-fig-1} illustrates the differences of using grisms and
prisms as dispersive elements by comparing the dispersion and the sensitivity
of the HRC grism G800L (left) and the HRC prism PR200L (right). The G800L has
an almost constant dispersion over the entire accessible wavelength range.
The dispersion of the PR200L increases drastically towards longer wavelengths,
and any value for the dispersion must be accompanied by the wavelength at which
it is specified (see Tab.\ \ref{spec_modes} and the ACS Instrument Handbook).
The different properties of the prisms and grisms require a flexible data
reduction software to be able to reduce both slitless grism and prism data.\\

Following the demise of STIS in August 2004 the interest in using ACS grisms
and prisms has increased substantially, since it is the only optical-UV
spectral capability aboard HST. As a result, around 10\% of all approved
orbits in Cycle 14 are devoted to slitless spectroscopy mode with ACS
(Macchetto et al.\ 2005).

\section{Slitless spectroscopy from space}
\begin{figure}
\plotone{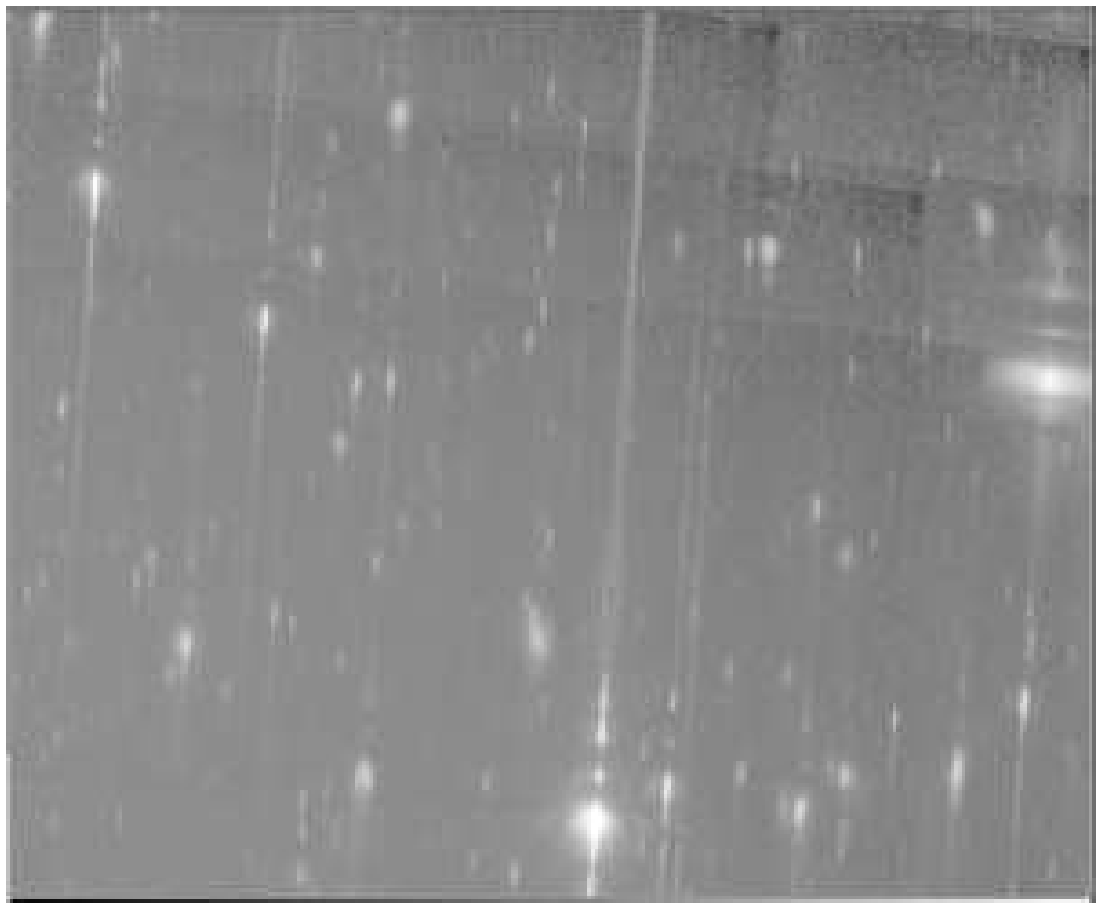}
\caption{A MultiDrizzle-combined WFC/G800L image. The large number of
spectra to be extracted from the data and the mutual contamination
of spectral orders are evident.}
\label{example-fig-2}
\end{figure}

Slitless spectroscopy with the ACS has some distinctive features which
clearly separate the technique from spectroscopy with slits.
Its advantages are:
\begin{itemize}
\item The sky background is extremely low. Typical background count rates for
the combinations WFC/G800L, HRC/G800L and HRC/PR200L are $0.1$, $0.006$ and
$0.04$\,e$^-$/s/pix, respectively. With typical exposure times in the range
$1000-2000$\,s the read-out noise of $\sim 4.9e^-$ remains an important
to even dominant source to the overall noise in the images.
\item In contrast to the ground-based sky background, which is usually
dominated by emission lines, the HST sky background is much smoother,
which makes the background removal less problematic. Since the removal
of a background with emission lines always leaves variations in the
signal-to-noise ratio and therefore leads selection effect in the analysis
of the data. ACS slitless spectroscopy avoids these selection effects.
\item ACS slitless spectroscopy is associated with an extremely large
data yield. To illustrate this Figure \ref{example-fig-2} displays a
MultiDrizzle-combined WFC/G800L image. While it is not general to extract
slitless spectra from MultiDrizzle-combined images such as
Fig.\ \ref{example-fig-2}, such coadded images give a first impression on 
what can be expected from the data. The image shows the spectra of thousands
of objects, which all can be extracted from the data. In ACS slitless
spectroscopy, the number of objects per field to be extracted is solely
determined by the depth of the observation.
\item Also the typical advantages of the HST, such as the compact
point spread function and the high stability of the instrument,
apply to ACS slitless spectroscopy.
\end{itemize}
The disadvantages of the ACS slitless spectroscopy modes are:
\begin{itemize}
\item The mutual contamination of spectra is an ubiquitous
phenomenon in slitless spectroscopy, especially since the absence of
slits or masks allows a contamination in the spectral direction
(see Fig.\ \ref{example-fig-2}). The contamination can occur even from
higher spectra orders. In Fig.\ \ref{example-fig-2} this becomes
evident when looking at the various spectral orders of the bright stars,
which cover a large area and of course overlay the spectra
of the fainter objects. Contamination affects all spectra to various
degrees, and the reduction software has to deal with contamination
(see Section \ref{axe_software} and Walsh, K\"ummel \& Larsen 2005).
\item The reduction of slitless spectroscopic data is quite different from the
usual extraction of spectra taken with slits. It requires different methods,
the application of different calibrations and the usage of different software.
As a consequence, the astronomer needs some time to get familiar with slitless
data and its reduction concepts.
\end{itemize}

\section{Reducing slitless spectroscopic data}
\label{red_slitless}
In 'traditional' spectroscopy with slits, the aperture (slit or mask) 
defines a framework for the trace definition and the wavelength solution.
The information derived from calibration data (flatfield- and
arc-exposures) taken with the identical instrumental setup is directly
transferred to the science data to extract the spectra.

\begin{figure}
%\epsscale{0.30}
\plotone{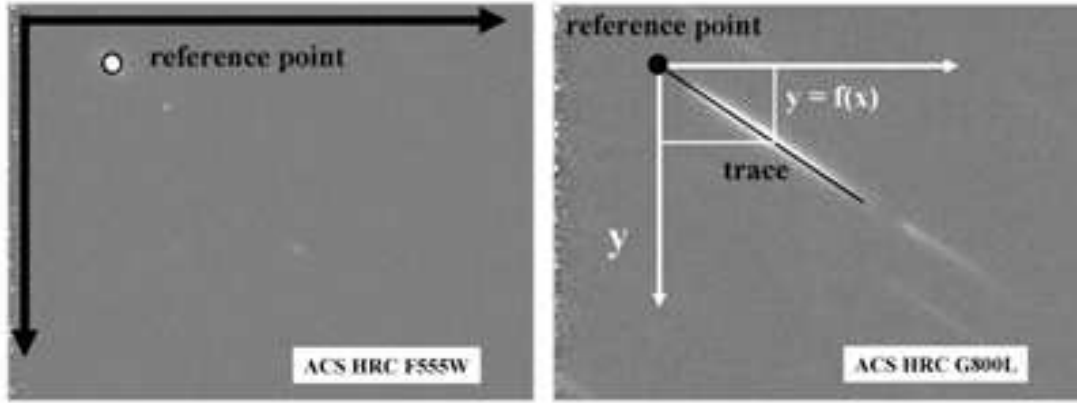}
\caption{The HRC/G800L grism image (right panel) and the associated direct
image taken with the F555W filter (left panel). The reference point for an
example object is determined with SExtractor in the F555W image and then
transferred to the G800L image. In the grism image the  reference
point is the origin of the coordinate system to describe the location
of the trace. Finally, the wavelength solution is applied as
a function of the trace length.}
\label{example-fig-3}
\end{figure}

In slitless spectroscopy, however, such a framework does not exist.
The exact location of the objects on the science data is unknown,
and it is impossible to establish a trace description and a wavelength
solution on the basis of slitless data alone such as in the right panel in
Figure \ref{example-fig-3}.
To base a framework or coordinate system for the trace description
and wavelength solution, a so-called {\it reference point} is needed for every
spectrum which is to be extracted. The reference point is the basis for the
origin of a local coordinate system (see Fig.\ \ref{example-fig-3},
right panel) to define the trace description and apply the wavelength
solution. In ACS slitless spectroscopy, the reference points of all objects
are established on a direct image such as shown in Fig.\ \ref{example-fig-3},
left panel, which is associated with every slitless image. Then the 
extraction procedure is a follows:
\begin{enumerate}
\item All object positions and therefore reference points are determined
on the direct image, which was taken very close on time to the slitless image.
This is done either directly with a source detection software such as
SExtractor, or indirectly by computing their pixel coordinates
on the direct image.
(left panel in Fig.\ \ref{example-fig-3}).
\item The reference positions are transferred to the slitless image
(right panel in Fig.\ \ref{example-fig-3}).
\item A so-called {\it aXe configuration file}, which was assembled using
calibration data (see next Section), gives the prescription to define
the trace description and the wavelength solution for
every reference position on the slitless image.
\item The object spectra are finally extracted from the slitless image.
\end{enumerate}

\section{The calibration of the slitless modes}
The main calibration products acquired for an ACS slitless mode are:
\begin{itemize}
\item The aXe configuration file to describe trace description 
and wavelength solution.
\item A sensitivity file for the flux calibration of every
spectral order.
\item A flatfield to be able to apply pixel-to-pixel gain corrections.
\end{itemize}
Since there are no on-board calibration lamps for ACS, suitable astronomical
targets with known fluxes and known spectral features
were observed in dedicated calibration proposals to establish the flux
calibration and wavelength calibration, respectively.

\begin{figure}
%\epsscale{0.30}
%\plotone{Pas_fig5.ps}
\includegraphics[angle=0,width=.5\textwidth]{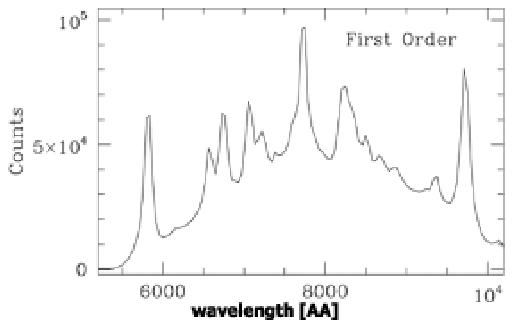}
\includegraphics[width=.5\textwidth]{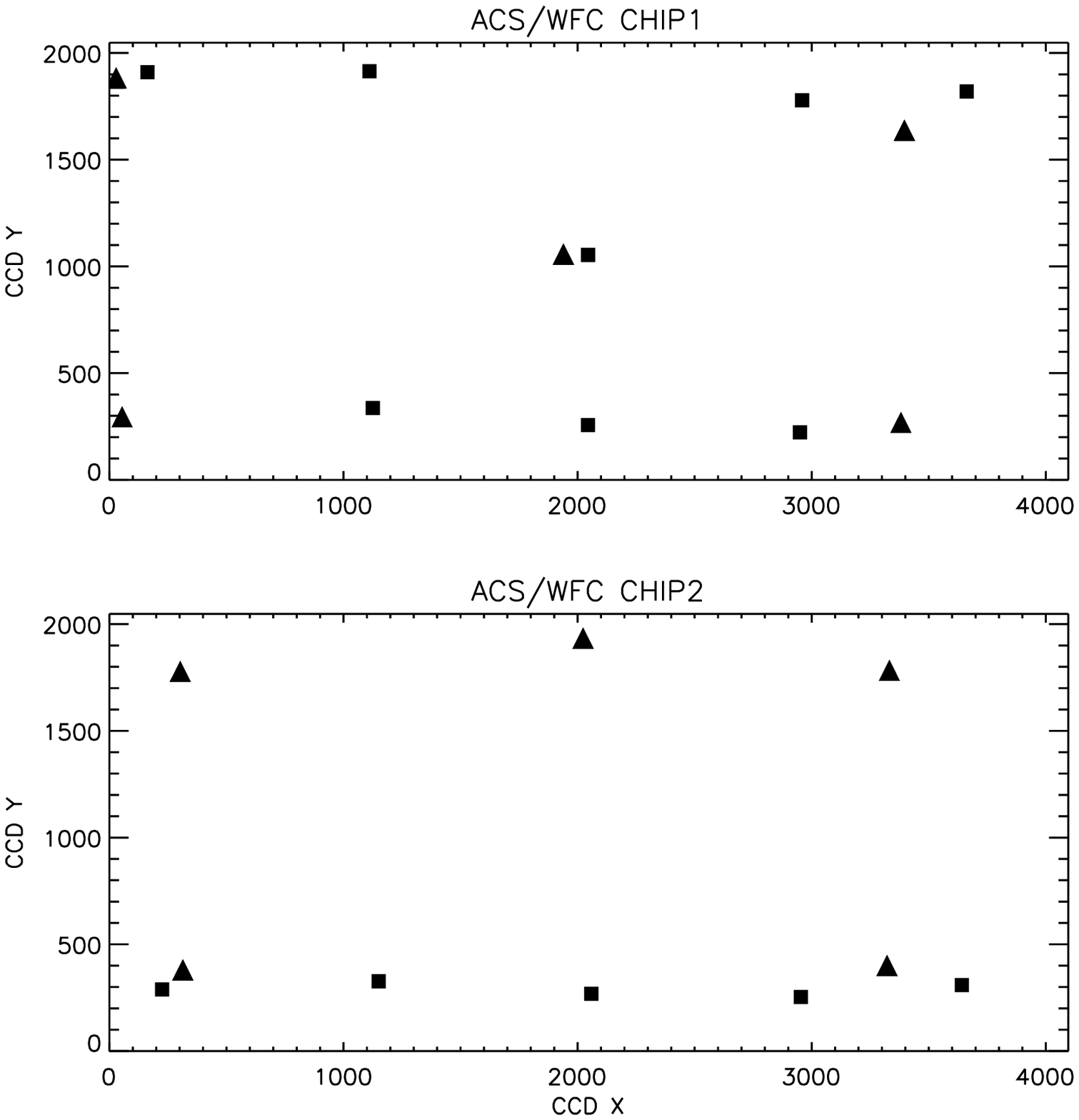}
\caption{Left panel: The spectrum of the Wolf-Rayet star W45, which was used
to establish a wavelength calibration for the WFC/G800L mode. Right panel:
The positions of astronomical targets across the WFC/G800L FoV to determine
the 2D-dependence of the calibrations.}
\label{example-fig-4}
\end{figure}

For the flux calibration and trace definition, white dwarf standards
could be used for all ACS slitless modes. For the wavelength calibration,
planetary nebulae and Wolf-Rayet stars were observed for the optical modes
WFC/G800L and HRC/G800L (Pasquali et al.\ 2005, Larsen \& Walsh 2005).
The left panel of Figure \ref{example-fig-4} shows the first order
spectrum of the Wolf-Rayet star WR45 as observed with the WFC/G800L.
The bright emission lines which spread over the entire spectral range
are identified to define the wavelength solution.
The ACS prism modes, which cover the near-UV (HRC/PR200L, SBC/PR120L and
HRC/PR110L), were wavelength calibrated using observations of planetary
nebulae and carefully redshift-selected QSO's
(Larsen, K\"ummel \& Walsh 2005).\\

In all ACS slitless modes, the trace definition, wavelength calibration and
sensitivity show variations across the FoV. For this reason
the astronomical calibration targets are observed at several positions
across the FoV. Figure \ref{example-fig-4} shows as an example the
positions of the
astronomical calibration targets across the WFC FoV. After establishing
the calibration at each position individually, a global 2D solution is
fitted to the individual solutions, which then enables to derive the
 calibration at every arbitrary position in the FoV.\\

The absence of on-board calibration lamps also inhibits the conventional
approach for flatfielding. But even if calibration lamps were present,
taking a flatfield exposure for every science exposure, as it is usually done
in slit spectroscopy, could not deliver a proper flatfield.
In ACS slitless spectroscopy the
objects and their exact position, and therefore the wavelength attributed to
the pixels, are unknown prior to the data reduction. Moreover there are
several wavelengths associated to each pixel in regions where the spectra
of multiple objects overlap, and this is not possible in the conventional
approach.

\begin{figure}
%\epsscale{0.30}
\plotone{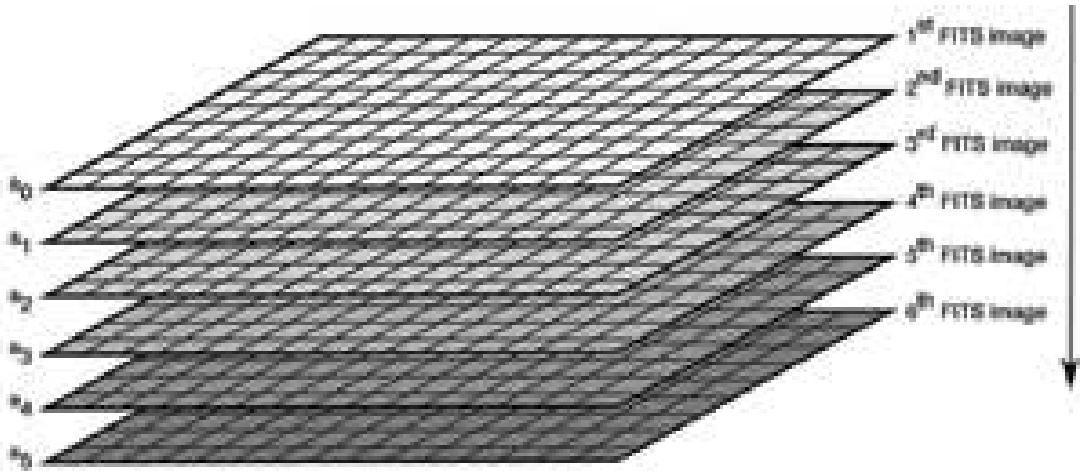}
\caption{The structure of a flatfield cube. Each extension of the fits file
gives the coefficients to compute the flatfield correction for every pixel.}
\label{example-fig-5}
\end{figure}

As a consequence, the flatfield used in ACS slitless spectroscopy 
must be able to correct any pixel for any wavelength. The solution
to this problem is a flatfield cube, as shown in Figure \ref{example-fig-5}.
A flatfield cube is a multi-extension fits file, and every extension contains
the coefficients to compute the gain correction for any given pixel
$(i,j)$ at any wavelength $\lambda$ according to the formula: 
\[
FF(i,j,\lambda) = a_0(i,j) + a_1(i,j)*\lambda + a_2(i,j)*\lambda ^2 
+ a_3(i,j)*\lambda ^3 \ldots
\]
The flatfield cubes for the different spectroscopic modes are created
from photometric flats, and further details on the flatfielding
is given in Walsh \& Pirzkal (2005).

\section{The aXe spectral extraction software}
\label{axe_software}
\begin{figure}
\epsscale{0.85}
\plotone{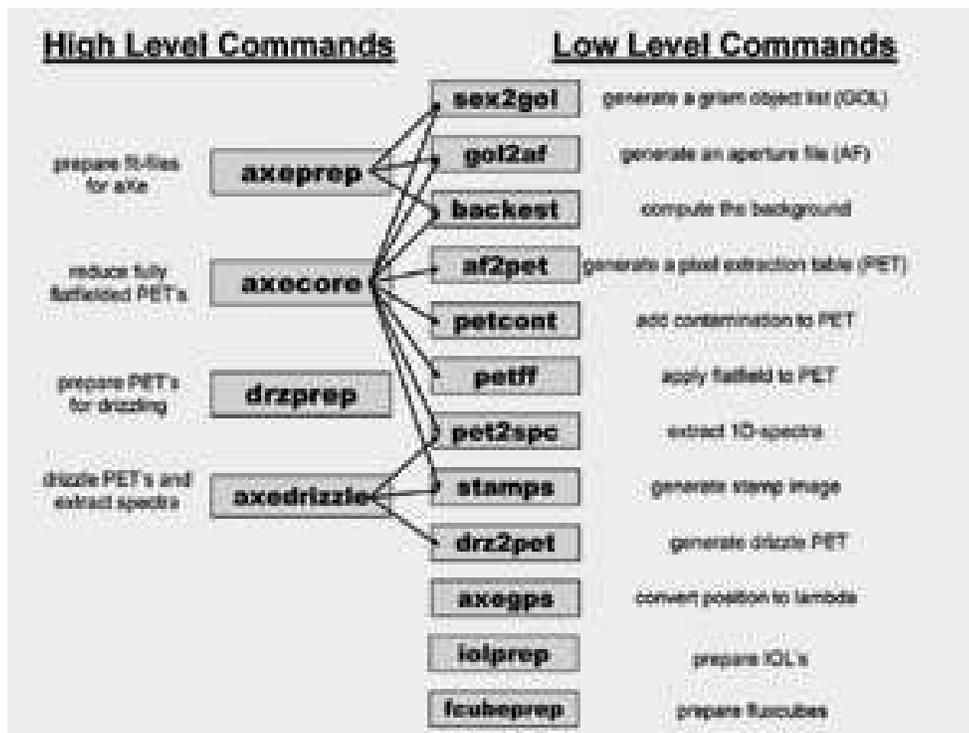}
\caption{The list of aXe tasks. The arrows indicate the interaction
between the High Level and the Low Level Tasks. The right column
describes the reduction step executed by a certain task.}
\label{example-fig-6}
\end{figure}
The aXe software package was specifically designed to extract spectra
in a consistent manner from all slitless spectroscopic modes of the ACS.
While the reduction of slitless spectroscopic ACS data was the driving force
behind the development of aXe, the software package was designed to allow
also the reduction of spectroscopic data from other instruments without need
for fundamental software changes, and currently an investigation
is done to apply aXe to data taken with the FORS2-MXU unit at the VLT
(see K\"uemmel et al.\ 2006 and Kuntschner et al.\ 2005).

aXe is a PyRAF/IRAF package with several tasks (see Figure \ref{example-fig-6})
which can successively be used to produce extracted spectra from
slitless data. As can be seen in Fig.\ \ref{example-fig-6}, there exist two
classes of aXe tasks:
\begin{enumerate}
\item The Low Level Tasks work on individual images. Their aim
is to perform a certain reduction step on a particular image.
\item The High Level Tasks work on data sets. Their aim is to do a sequence
of processing steps on several images.
\end{enumerate}
Often the High Level Tasks call Low Level Tasks to perform a certain
reduction step on individual images, and Fig.\ \ref{example-fig-6}
shows their interaction. The High Level Tasks were designed
to cover all steps of the aXe data reduction, and working with aXe usually
means to apply the High Level Tasks to data sets. Due to the large data
yield in ACS slitless data, which makes it impossible to extract
every spectrum individually, aXe is built as a semi-automatic data reduction
system. After the object positions are determined on the direct images
(see Sect.\ \ref{red_slitless}), aXe runs automatically and creates
besides the final, extracted spectra additional intermediate products,
such as 2-dimensional stamp images of the spectra. The intermediate
data products help the user to check the reduction procedure and to
fine-tune the extraction parameters.

The aXe tasks are implemented in Python. To work on the pixel data
(such as flatfielding or extracting 1D spectra), which
requires a large computational speed, the Low Level Tasks call executables
which are implemented in standard {\bf C}. Via PyRAF the aXe tasks are fully
embedded into the STSDAS software package, and aXe users do not have to
leave their familiar data reduction environment to work with ACS slitless
data. The aXe package evolves continuously, and together with STSDAS
new aXe versions are released about once a year. The current aXe-1.5 was
released together with STSDAS 3.4 in November 2005. Between these large
releases there are occasional smaller software releases via the aXe
webpage (http://www.stecf.org/software/aXe/) to publish bug-fixes
or together with new aXe configuration files.

The two main drivers behind the development of the aXe software package are:
\begin{enumerate}
\item Improvements for the user convenience.
\item Adding new functionality to aXe.
\end{enumerate}
The former was the motivation for developing the High Level Tasks
in aXe-1.4, which greatly reduced the number of commands with which
the user has to become acquainted.
In aXe-1.5 the task {\tt iolprep} was added, which
is a new tool to generate SExtractor lists with object positions for a set
of slitless images in a standard scenario.

As new functionality {\it optimal weighting} and the so-called
{\it quantitative contamination} were added
in aXe-1.5. The old aXe method of recording contamination associates to every
pixel in the slitless image the number of spectra of which the pixel
is a part.
This information is processed in the 1D spectral extraction, and each spectral
element has as contamination information the number of other spectra its
input pixels contributed to.
This method of recording contamination gives no clue
on how much the contaminating objects actually influence the extracted
object flux.

In quantitative contamination the contributing flux from other
objects to every spectral bin is estimated according to a simple emission
model. The quantitative contamination estimate is a very good tool for the
user to decide which data points he can trust. A detailed description
of the quantitative contamination method and of the optimal weighting
implemented in aXe-1.5 is given in Walsh, K\"ummel \& Larsen
(2005).\\

\begin{figure}
\epsscale{1.0}
\plotone{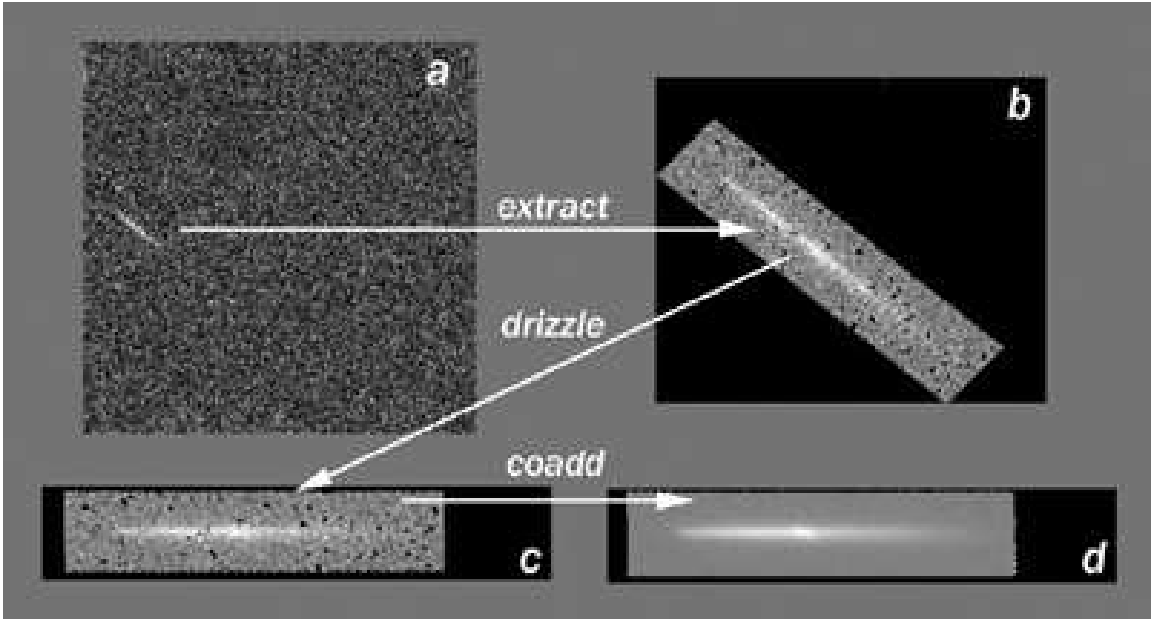}
\caption{The aXedrizzle method of combining 2D spectra. The object in panel
a is extracted as a stamp image (b), which is drizzled to an image with
constant dispersion and constant pixel scale in spatial direction. The
1D spectrum is finally extracted from the deep, coadded 2D spectrum (d).}
\label{example-fig-7}
\end{figure}
Another important addition to the functionality of aXe is the {\it axedrizzle}
method, which was first released in aXe-1.4 (K\"ummel et al. 2005).
With the aXedrizzle method the individual 2D spectra of an object
(see Figure \ref{example-fig-7}a, b)
are coadded to a single, deep 2D image (Fig.\ \ref{example-fig-7}d) before
extracting the 1D spectrum. The combination of the individual 2D spectra is
done with the ``Drizzle'' software (Fruchter \& Hook 2002), which is well known
from HST imaging. This method of combining the data has several advantages:
\begin{itemize}
\item Resampling to a uniform wavelength scale and an orthogonal spatial
direction with a constant pixel scale is achieved in a single step.
\item Pixel weighting is handled correctly.
\item The coadded 2D spectrum can be viewed to detect problems.
\end{itemize}
Due to the highly non-linear form of the prism dispersion
(see Fig.\ \ref{example-fig-1}), the aXedrizzle method is restricted to
the ACS grism modes WFC/G800L and HRC/G800L only.

\section{The aXe visualisation of spectra}
\begin{figure}
\epsscale{0.9}
\plotone{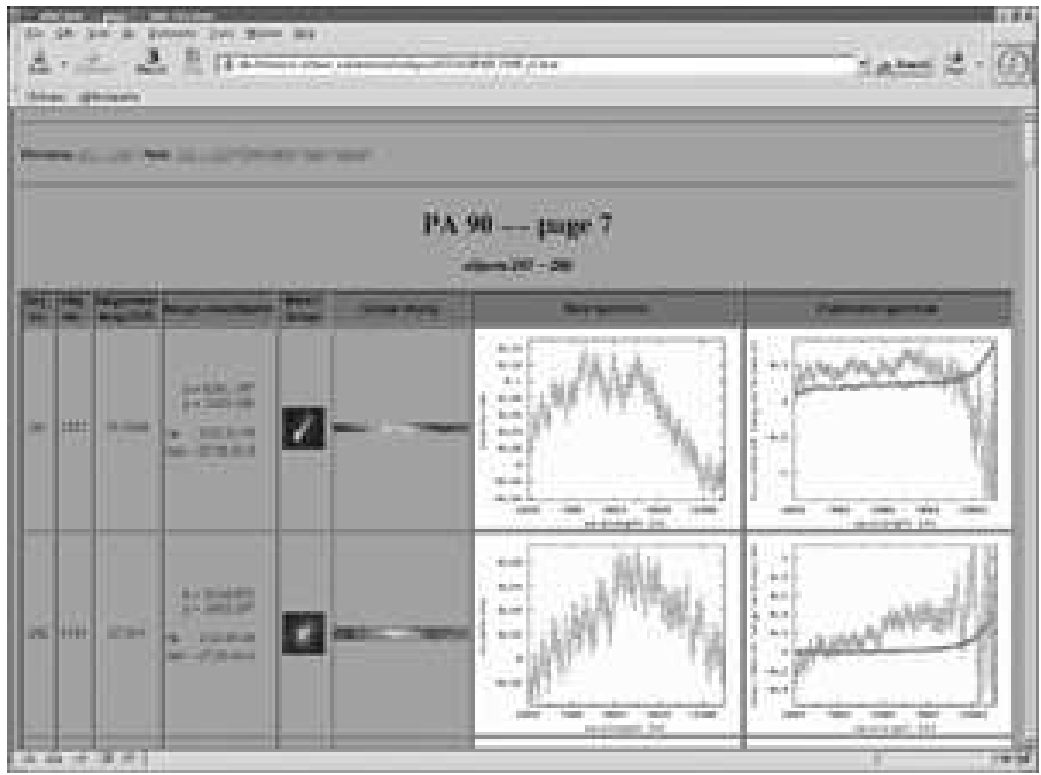}
\caption{A webpage created by aXe2web. The various rows contain
information on individual objects such as their reference number, magnitude,
position, stamp images and spectra}
\label{example-fig-8}
\end{figure}
As indicated above, a data set with ACS slitless images may contain hundreds
or even thousands of spectra, and a visual inspection of each individual
spectrum is very tedious. To help the user digest the large amount of data,
the tool aXe2web was implemented, which produces browsable webpages
for a fast and discerning examination of spectra (Walsh \& K\"ummel 2004).

aXe2web uses the direct image, the SExtractor catalogue, the aXe stamp images
and the extracted spectra to produce an HTML summary containing a variety
of information for each spectrum. Figure \ref{example-fig-8} shows part of
an HTML page produced by aXe2web. Each object produces a line in the HTML
page which lists the sequence number,
the reference number, the magnitude of the direct image object, the
coordinates (image and RA/DEC) of the direct image object, the direct
stamp image, the grism/prism stamp image and the 1D extracted spectrum
in $counts/s$ and flux.

To facilitate an easy navigation within a data set, an overview and an
index page accompany the object pages (Fig.\ \ref{example-fig-8}) which show
the detailed object information. The overview page contains for each
object the basic information, e.g. object positions and magnitudes,
and the index page includes a table with the ordered reference
numbers of all objects. Direct links guide from the overview page and
the index page to the corresponding location on the object pages.

\section{Conclusions}
\begin{itemize}
\item All aspects of ACS slitless spectroscopy (calibration, software etc.,)
are supported such that users can obtain and reduce slitless data in 
a pipeline way. 
\item ACS slitless spectroscopy is successfully used in various science
projects such as the HUDF HRC Parallels
(Walsh, K\"ummel \& Larsen 2004), high redshift supernovae research
(Riess et al. 2004) and the search for high redshift galaxies
(the GRAPES and PEARS programmes, see Pirzkal et al., 2004).
\item More information about ACS slitless spectroscopy, the calibration
and the aXe software is given on the aXe webpages
at http://www.stecf.org/software/aXe/
\item User support concerning all topics related to ACS slitless
spectroscopy is provided by the ACS group at the Space Telescope - European
Coordinating Facility (ST-ECF). The centralised email address for requests is
acsdesk@eso.org.
\end{itemize}
%-----------------------------------------------------------------------
%			      References

\end{document}